\documentstyle[aps,prd,multicol]{revtex}
\tighten

\begin{document}
\title{Single neutral heavy lepton production at electron-muon colliders.}
\author{F.M.L. Almeida Jr.\thanks{E-mail: marroqui@if.ufrj.br},Y. A. 
Coutinho\thanks{E-mail: yara@if.ufrj.br}, \\
J. A. Martins Sim\~oes\thanks{E-mail: simoes@.if.ufrj.br}, 
M.A.B. do Vale\thanks{E-mail: aline@if.ufrj.br}\\
Instituto de F\'\i sica\\
Universidade Federal do Rio de Janeiro, RJ, Brazil \\}

\maketitle
\begin{abstract}
\par
New heavy Majorana and Dirac neutrinos production at future electron-muon colliders are investigated. The production of a single heavy neutrino is shown to be more relevant than pair production when comparing cross sections and neutrino mass ranges. The process $ e^\pm \mu^\mp \longrightarrow {\nu} \, \ell^{\pm}\, W^{\mp}$ is studied including on-shell and off-shell heavy neutrino effects. Distributions are calculated including hadronization effects and experimental cuts that suppress background, in order to have a clear signal for heavy neutral leptons.
\end{abstract}
\par

\begin{multicols}{2}

\par
There is an increasing experimental evidence for neutrino oscillations and non zero neutrino masses\cite{PDG}. The smallness of neutrino masses is generally understood as a consequence of some see-saw mechanism. This brings the question of the possibility of new heavy neutrino states. So far none of these new states was experimentally observed\cite{PDG,ZUB} with masses up to $M_N\simeq$ 100 GeV. For higher masses there are many suggestions of experimental possibilities in the next high energy hadron-hadron colliders\cite{YSP}, in electron-positron linear accelerators\cite{ARS,DJO,ACV}, in neutrinoless double-beta decay\cite{ROD}. The properties of these new heavy states are a central point in many of the theoretical models proposed as extentions of the present standard model of elementary particle physics. Besides the masses and mixing angle values, an important point is the Majorana or Dirac nature of new heavy states. This is directly connected with lepton number conservation and to the general symmetries of any extended model. 
\par
The suggestion of a new type of electron-muon colliders\cite{CHO} seems to be an interesting proposal  to verify the properties of new heavy leptons. Muon beams are well known to have a reduced synchrotron radiation loss and the absence of an Z-mediated $s$ channel makes cleaner the high energy properties of charged current interactions. Another interesting point is the fact that $ e \mu $ colliders will test directly the properties of two leptonic families. This option was considered recently by Cveti\u c and Kim \cite{CVE}, who studied the production of a pair of heavy Majorana leptons in the reaction $ \ell^- \ell'^+ \longrightarrow N N \longrightarrow W^{\pm}\ell_i^{\mp}W^{\pm}\ell_j^{\mp} $ for N on and off shell.
\par
In this paper we call attention to a more feasible channel for detecting neutral heavy leptons in $ e \mu$ colliders. In many models the mixing of neutral and heavy neutrino states implies \cite{ARS,DJO} that   in the light-to-heavy neutrino vertex we have a single power of the mixing angle, contrary to the double mixing angle power in the heavy-to-heavy neutrino vertex. Another factor that favors single heavy lepton production is a smaller phase space suppression.  We are then lead to study the channel  $e^\pm \mu^\mp \longrightarrow {\nu} \, \ell^{\pm}\, W^{\mp}$. We have calculated the standard model background contribution for this process and shown that, with appropriated cuts, it can be reduced well bellow the signal. In order to have a more realistic estimate for the signal and background distributions we have hadronized the final W. We discuss throughout this paper both the Majorana and Dirac heavy neutrino contributions.
\par

Most extended models predict new fermions and new gauge vector bosons. Since we presently have no signal for new interactions, we will make the hypothesis that the new heavy neutrino states behave as $SU_L(2)\otimes U_Y(1)$ basic representations. We can resume the new particle  interactions in the neutral and charged current lagrangians:

\begin{equation}
{\cal L}_{nc}=-\frac{g}{4c_W } {sin{\theta_{mix}}} Z_{\mu}\overline{\psi_N}\gamma^{\mu}
\left(1-\gamma_{5}\right)\psi_{\nu} + h.c..
\end{equation}

and

\begin{equation}
{\cal L}_{cc}=-\frac{g}{2\sqrt{2}} {sin{\theta_{mix}}}
W_{\mu}\overline{\psi_N}\gamma^{\mu}\left(1-\gamma_5\right)\psi_e + h.c.
\end{equation}
and similar terms for the other leptonic families. In the Majorana case, each completely neutral heavy lepton is coupled to all charged leptons and we are considering only the lower mass state. For the Dirac case, in principle we have three different heavy neutrinos, one for each family. In this last case we have considered their masses to be all equal.
\par
 Throughout this paper we will suppose that mixing angles for heavy-to-light neutrinos and new heavy neutrino masses are independent parameters \cite{BUC}. The light neutrinos couplings to the neutral Z are given by $ g_{V,A}=g_{V,A}^{SM} - {sin^2{\theta_{mix}}/2}$. Taking the experimental values for $g_{V,A}$ and the standard model predictions we obtain a small upper bound for $\theta_{mix}$.  A recent estimate \cite{YPS} gives $\sin^2 \theta_{mix} < 0.0052$ with $95\%$ C.L. This limit value is used throughout this paper for all curves and distributions.

\par
The decay modes for these leptons, in the Majorana case \cite{YSP} must include both signatures $N \longrightarrow \ell^{\mp}\, W^{\pm}$  and $ N \longrightarrow \nu_{\ell}\,(\bar\nu_{\ell})\, Z$, with $\ell=e,\mu,\tau$. For the Dirac case, we have the simple decays $ N_{\mu} \longrightarrow \mu^-\, W^+ $ and $ N_{\mu} \longrightarrow \nu_{\mu}\, Z$. For the specific channel considered in this paper we have an undetected final neutrino state. In the heavy Dirac neutrino production we have lepton number conservation, and only reactions like $ e^- \mu^+ \longrightarrow \nu_{e} \, \mu^+\, W^-$ are allowed. But for the Majorana case, as we have lepton number violation, we must sum over all the final neutrino and antineutrino states in $ e^- \mu^+ \longrightarrow  \mu^+\, W^-\, \nu_{i} (\bar\nu_{j})$. 
\par
From equations 1 and 2 we can readily calculate the decay widths for the neutral leptons and find

\begin{eqnarray}
\Gamma(N \rightarrow \nu_i Z)={ g^2 \sin^2\theta_{mix}M_N \over{128 \pi c_W^2}} f(y_Z)\nonumber\\
\Gamma(N \rightarrow \ell^\pm W^\mp)={ g^2 \sin^2\theta_{mix} M_N \over{64 \pi}} f(y_W)
\end{eqnarray}
\bigskip
\noindent 
where $ f(y)=y-3/y+2/{y^2} $ and $ y=(M_N/M_{Z,W})^2$.
\par

In Fig. 1 we show the structure of the Feynman diagrams that contribute for the signal. We have included in our results all the off-shell and on-shell contributions. The calculations for both the standard model background and signal were done using the high energy program CompHep \cite{HEP}. We call attention to the fact that the N width is very narrow for lower heavy neutrino masses and increase for higher masses. Propagator effects must then be handled with  care. Another possible background contribution comes from $ e^- \mu^+ \longrightarrow e^- \, \mu^+\, Z $ where the final electron goes along the initial electron direction and escapes from the detection. This contribution is almost eliminated by requiring a good resolution in the hadronization, with an invariant hadronic mass peaked around $M_W$.

\par
The cross sections for single and pair heavy neutrino production (on-shell) are displayed in Fig. 2 for $ \sqrt{s}=$ 2000 GeV. In all our results  we have done general detector cuts
$E_{lepton}> 5$ GeV and $ -0.995< cos {\theta_i}<0.995 $, where all angles are defined relative to the initial electron. In the Majorana case we have summed over all the possibilities of undetected final neutrino states. This figure  shows clearly the dominance of single heavy neutrino production over pair production. We also show the possibility of Majorana neutrino contribution to $ e^- \mu^+ \longrightarrow W^- W^+$ according to Fig. 1b. This channel dominates single and pair Majorana production for neutrino masses  above $\sqrt s$. In Fig. 3 we have  the total cross sections for the signal (Dirac and Majorana) and standard model background for the specific final states $ e^- \mu^+ \longrightarrow invisible \, \mu^+\, W^-\,$, where $ invisible $ means the  sum of all undetected possible final neutrino and anti-neutrino states. The standard model background is clearly above the signal and the central problem for a possible experimental detection of heavy neutral leptons is to study the distributions that could separate signal from background.
\par
Angular distributions can be very helpful in separating signal from background as well as Majorana from Dirac heavy neutrinos. In Fig. 4 we show the angular distribution ($cos \theta_{\mu}$) for the final muon (relative to the initial electron). The standard model distribution is peaked at $cos\theta_{\mu}\simeq -1$. For lower heavy neutrino masses we have different Dirac and Majorana distribution shapes but for higher masses this difference almost disappears. In Fig. 5 we show the distribution of the angle between the final muon with respect to the  direction of the heavy lepton in its center of mass ($cos \theta^*_{\mu}$). This distribution shows that we can eliminate most of the standard model contribution with a cut like $ cos {\theta^*_{\mu}} < 0.5 $ and that this cut is almost independent of the heavy neutrino mass.
\par
In order to enhance the relation signal to background, we can apply some heavy neutrino mass dependent cuts, based on kinematical relations only. Let us consider the case where at $\sqrt{s}$ center of mass energy particles A and B are produced and then
the particle B decays into particles C and D. Let us suppose further that the particles A and C are massless. After doing a simple calculation we obtain that the energy of the particle C, $E_C$, should be in the range: 

\begin{equation}
\frac{1}{2} \frac{\sqrt{s}(M_{B}^2-M_{D}^2)}{M^2_{B}} <E_C< 
{1\over 2} {(M_{B}^2-M_{D}^2)\over \sqrt{s}}
\end{equation}

The shaded area in Fig. 6 shows the allowed region for the muon energy $E_{\mu}$ when $\sqrt{s}=$ 2000 GeV and final state $\nu$, $\mu^+$, $W^-$. For $M_N=200$ GeV, 800 GeV and 1600 GeV, we have 8.38 GeV $<E_{\mu} < 838.62$ GeV, 158.38 GeV $< E_{\mu} < 989.91$ GeV and 638.38 GeV $<E_{\mu} < 997.47$ GeV, respectively.

Another interesting heavy neutrino mass dependent variable is the angle between the final particles C and D, originated from the particle B decay in the center of mass of the incident particles.  After doing a Lorentz transformation, we obtain that $cos\theta_{CD}$ should be in the region:

\begin{equation}
 -1< cos\theta_{CD}<\frac{\gamma_v^2\sqrt{E^{*2}_C+M_D^2}-E^{*}_C}{\sqrt{1+\gamma_v}\sqrt{\gamma_v^2 (E^{*2}_C+M_D^2)
+E^{*2}_C}}
\end{equation}

where

\begin{equation}
\gamma_v=\frac{(s-M_B^2)}{2 \sqrt{s}M_B}$$\\
\end{equation}

and

\begin{equation}
E_C^{*}=\frac{(M_B^2-M_D^2)}{2 M_B}
\end{equation}

The shaded area on Fig. 7 shows the allowed kinematical region as a function of the heavy neutral particle mass for our case where $\sqrt{s}=$ 2000 GeV and A, B, C and D are the standard neutrino, heavy neutral particle, muon and gauge boson W, respectively. For $M_N=$ 200 GeV, 800 GeV and 1600 GeV, we have for $ \cos\theta_{W \mu}$ the following upper limits: $0.914$, $0.0588$ and $-0.9003$, respectively. Dynamical bounds will further limit this area. Another important kinematical constraint comes from the fact that on-shell heavy neutrino contributions are dominant.  Then the missing energy of the final neutrino has a sharp distribution given by 
\begin {equation}
 E_{\nu}=( s-M_N^2)/2\sqrt{s}
\end{equation}
\par
The most definite separation between signal and background is given by the correlation between the final missing energy and the invariant visible mass of the system {\it muon} + {\it hadrons}. For the signal the missing neutrino energy is strongly peaked by Eq. 9 and the invariant visible  {\it muon} + {\it hadrons} mass is also peaked around $M_N$.  In order to have a more realistic estimate for the experimental possibilities of detection of heavy neutrino effects we have hadronized the final W, for the signal and the background, using the program Pythia \cite{PIT}. The results are shown in Figs. 8, 9, 10 for heavy Majorana neutrino masses of 200, 800 and 1600 GeV respectively. In Fig. 8a we show the arbitrary number of events plot in the plane $E\!\!\!\slash $ and {\it muon} + {\it hadrons} invariant mass. For $M_N=$ 200 GeV we include only the general detector cuts, $E\!\!\!\slash > 5$ GeV; $E_\mu > 5$ GeV; $| \cos\theta_\mu | < 0.995$; | $\cos \theta_{hadron} | < 0.995$,  and require an hadronic invariant mass limited to $ M_W \pm 2$ GeV. In this heavy neutrino mass region the background is clearly mixed with the signal. If we start doing more restrictive cuts we can reduce the background contribution. In Fig. 8b we have  done the cut $E\!\!\!\slash  < 1050$ GeV, according to Eq. 9, and added the mass independent cut $cos \theta_{\mu}^* < 0.5 $.  If we finally make the more restrictive mass dependent cuts $ 0.85 < cos \theta_{W \mu} < 0.95$ and $E_{\mu} < 800$ GeV we arrive at the result shown in Fig. 8c, where the background was substantially reduced. A similar procedure can be done for higher masses. For instance, in Fig. 9a, even with the general detector cuts we can see the signal above the background. If we improve our cuts to the mass independent region $cos \theta_{\mu}^* < 0.5 $, use $E\!\!\!\slash  < 1050$ GeV, and employ  the more mass dependent cuts $cos \theta_{W \mu} < 0.05 $ and  158 GeV $< E_{\mu} < 990$ GeV, we have practically only the signal in Fig. 9b. The same situation is found for $M_N=$1600 GeV. In Fig. 10a, the signal is in a region of very low background. If we improve  our general and mass independent cuts with new $cos \theta_{W \mu} < -0.9 $ and  638 GeV $< E_{\mu} < 997$ GeV, only the signal is left unaffected, as displayed in Fig. 10b. 

\par
For single heavy neutrino production, the relation signal to background increases as the heavy neutrino mass increases. Our analysis of single heavy neutrino production via  the process $ e^- \mu^+ \longrightarrow  invisible \, \mu^+\, W^-$ can be generalized for any process of the type $ e^\pm \mu^\mp \longrightarrow  invisible  \, \, \ell^{\pm} \, hadrons $. For values of the center of mass energies other than $\sqrt s=$ 2000 GeV, we have similar conclusions: heavy neutrino masses can be investigated up to $\sqrt s$; angular and kinematical cuts can help in establishing distributions that can differentiate signal from background and the strong correlation between  the variables $E\!\!\!\slash$ and $ M_{\mu + hadrons}$ gives a very clear signature for the signal. 

\vskip 2cm
{\it Acknowledgments:} This work was partially supported by the
following Brazilian agencies: CNPq, FUJB, FAPERJ and FINEP.


\vskip 1cm
\vspace{1cm}
\LARGE
Figure Captions
\normalsize
\begin{enumerate}
\item General Feynman graphs for heavy Majorana and Dirac neutrino contribution to $e^-\mu^+ \longrightarrow {\nu}\, \mu^+ \, W^-$.
\item Single and pair production of on-shell heavy Dirac and Majorana neutrinos at $\sqrt s=$ 2000 GeV for $e^-\mu^+$ colliders ($\sin^2 \theta_{mix}=0.0052$).
\item Signal and background (standard model) contributions to $ e^- \mu^+ \longrightarrow 
invisible \, \mu^+\, W^-$ at $\sqrt s=$ 2000 GeV.
\item Final muon angular distribution relative to the initial electron $cos \theta_{\mu}$, for the standard model background and for Dirac and Majorana heavy neutrinos, at $\sqrt s=$ 2000 GeV.
\item Final muon angular distribution relative to the direction of the heavy lepton in its center of mass $cos \theta_{\mu}^*$, for the standard model background and for Dirac and Majorana heavy neutrinos, at $ \sqrt s=$ 2000 GeV.

\item Kinematical limits for the muon energy $E_{\mu}$ versus heavy neutrino mass $M_N$ at  $\sqrt s=$ 2000  GeV.
\item Angular distribution for the angle between the final muon and W, $cos \theta_{W \mu}$ at 
$\sqrt s=$ 2000  GeV.

\item Invariant visible mass $M_{ \mu \, + hadrons}$ and  missing energy $E\!\!\!\slash$  for background and signal for $M_N=$ 200 GeV (in arbitrary units). Fig. 8a was done with the general cuts, Fig. 8b was done with the additional mass independent cuts as discussed in the text and Fig. 8c with the more restrictive cuts.
\item  Same as Fig. 8, for $ M_N=$ 800 GeV at $ \sqrt s=$ 2000 GeV.
\item  Same as Fig. 8, for $ M_N=$ 1600 GeV at $ \sqrt s=$ 2000 GeV.

\end{enumerate}

\vfill\eject
\end{multicols}
\end{document}